\documentclass{article}

\usepackage{spconf}
\usepackage{times}
\usepackage{epsfig}
\usepackage{graphicx}
\usepackage{amsmath}
\usepackage{amssymb}
\usepackage{lipsum}  
\usepackage[comma,sort&compress,numbers]{natbib}
\usepackage[small,hang]{caption}
\usepackage{makecell}
\usepackage{multirow}
\usepackage{booktabs}
\usepackage{tabularx}
\usepackage[dvipsnames]{xcolor}
\usepackage{enumitem}
\usepackage{subfig}
\usepackage{float}
\usepackage{array}

\newcolumntype{P}[1]{>{\centering\arraybackslash}p{#1}}
\newcolumntype{M}[1]{>{\centering\arraybackslash}m{#1}}
\usepackage[breaklinks,colorlinks]{hyperref}
\usepackage{tabularx}
\usepackage{flushend}
\begin{document}
\title{A Subjective Quality Study for Video Frame Interpolation}
\name{Duolikun Danier\thanks{The authors acknowledge the funding from China Scholarship Council, the University of Bristol and the
MyWorld Strength in Places Programme.}, Fan Zhang and David Bull}
\address{Visual Information Laboratory, University of Bristol, One Cathedral Square, BS1 5DD, United Kingdom\\
\{Duolikun.Danier, Fan.Zhang, Dave.Bull\}@bristol.ac.uk}

\maketitle
\begin{abstract}
Video frame interpolation (VFI) is one of the fundamental research areas in video processing and there has been extensive research on novel and enhanced interpolation algorithms. The same is not true for quality assessment of the interpolated content. In this paper, we describe a subjective quality study for VFI based on a newly developed video database, BVI-VFI. BVI-VFI contains 36 reference sequences at three different frame rates and 180 distorted videos generated using five conventional and learning based VFI algorithms. Subjective opinion scores have been collected from 60 human participants, and then employed to evaluate eight popular quality metrics, including PSNR, SSIM and LPIPS which are all commonly used for assessing VFI methods. The results indicate that none of these metrics provide acceptable correlation with the perceived quality on interpolated content, with the best-performing metric, LPIPS, offering a SROCC value below 0.6. Our findings show that there is an urgent need to develop a bespoke perceptual quality metric for VFI. The BVI-VFI dataset is publicly available and can be accessed at \url{https://danier97.github.io/BVI-VFI/}.
\end{abstract}
\section{Introduction}
\label{sec:intro}

Video frame interpolation (VFI) has attracted increasing attention in the research community over the past few years~\cite{liu2017video,xu2019quadratic, danier2021spatio}. By generating non-existent frames between two consecutive original frames in a video with a relatively low frame rate~\cite{mackin2015study}, VFI up-converts the temporal resolution of a video, increasing the motion consistency and overall perceptual quality. VFI methods are used for the generation of slow motion content~\cite{jiang2018super}, and also serves as a useful tool for video compression~\cite{lu2016novel, ding2021neural}. 

In recent years, deep learning (DL) techniques, in particular using convolutional neural networks (CNNs), have stimulated the development of new VFI algorithms. These exist as two main paradigms: flow-based and kernel-based. Flow-based approaches~\cite{liu2017video, xu2019quadratic} focus on improving the accuracy of estimated optical flows that are used to warp the consecutive original frames. In contrast, kernel-based methods~\cite{niklaus2017video, lee2020adacof} learn adaptive local kernels to synthesise the output pixels. Several methods~\cite{bao2019memc, danier2021spatio} have also been developed to combine flow-based warping and kernel-based synthesis to achieve improved interpolation performance. 

Although significant progress has been achieved using these new VFI methods, there is very little research reported on the quality assessment of interpolated content. The most commonly used quality metrics are PSNR and SSIM~\cite{wang2004image}. In order to better predict the visual quality of interpolated videos, perceptual quality metrics including LPIPS~\cite{zhang2018unreasonable}, FRQM~\cite{zhang2017frame}, ST-GREED~\cite{madhusudana2021st},  VIF~\cite{sheikh2005information} and VMAF~\cite{li2016toward}, which were designed for various application scenarios, can also be employed. However, none of these methods have been fully evaluated on frame interpolated videos, and their correlation with subjective quality for this type of content is unknown. It is also apparent that, in order to evaluate the performance of these quality metrics, there is a requirement for video quality databases containing diverse content generated by various VFI algorithms. Such a database is absent in the existing literature.

In this context, a subjective study has been conducted based on a new database containing interpolated video content. This database, BVI-VFI, contains 36 reference sequences (from 12 original sources) with a spatial resolution of 1080p (HD) at three different frame rates: 30, 60 and 120fps. Each of these 36 reference videos is then down-sampled and interpolated by five VFI methods (two conventional and three DL based) to produce a total of 180 distorted videos. The subjective experiment employed a double stimulus test methodology to collect quality opinion scores of these distorted sequences. The video database and the associated ground-truth quality scores were then employed to evaluate eight popular quality metrics, including those most commonly used (PSNR, SSIM and LPIPS) in the VFI literature. The results show that none of these metrics provide satisfactory correlation with perceptual quality, which raises concerns when these assessment methods are used to compare VFI algorithms. As far as we are aware, ours is the first video quality database developed specifically for VFI, and it offers a valuable platform for developing and evaluating bespoke VFI quality metrics. 

The rest of this paper is organised as follows. Section~\ref{sec:database} details the development of the BVI-VFI database, and the methodology and configuration of the subjective experiment is described in Section~\ref{sec:experiment}. We then present the results and analysis in Section~\ref{sec:results}, and conclude the paper in Section~\ref{sec:conclusion}.

\section{The BVI-VFI Database}\label{sec:database}

This section presents the methodology used to obtain the 36 reference and 180 distorted sequences in the BVI-VFI database.

\subsection{Reference Sequence Selection}

Since the original video frame rate is an important factor affecting the difficulty of VFI~\cite{choi2020channel}, it is desirable to have frame rate as a controlled variable when evaluating VFI methods. Based on this consideration, we selected our source sequences from the BVI-HFR~\cite{mackin2018study} dataset which contains HD (1920$\times$1080p) video sequences at various frame rates up to 120fps and with diverse motion types. Twenty-two source videos at 120fps (the acquisition frame rate) were included in the initial selection pool, and then truncated to five seconds\footnote{This is to reduce the total experiment length based on the recommendation in \cite{moss2015optimal}.}.

\begin{figure*}[t]
\renewcommand*\thesubfigure{\arabic{subfigure}} 
    \centering
    \subfloat[Bobblehead]{\includegraphics[width=0.160\linewidth]{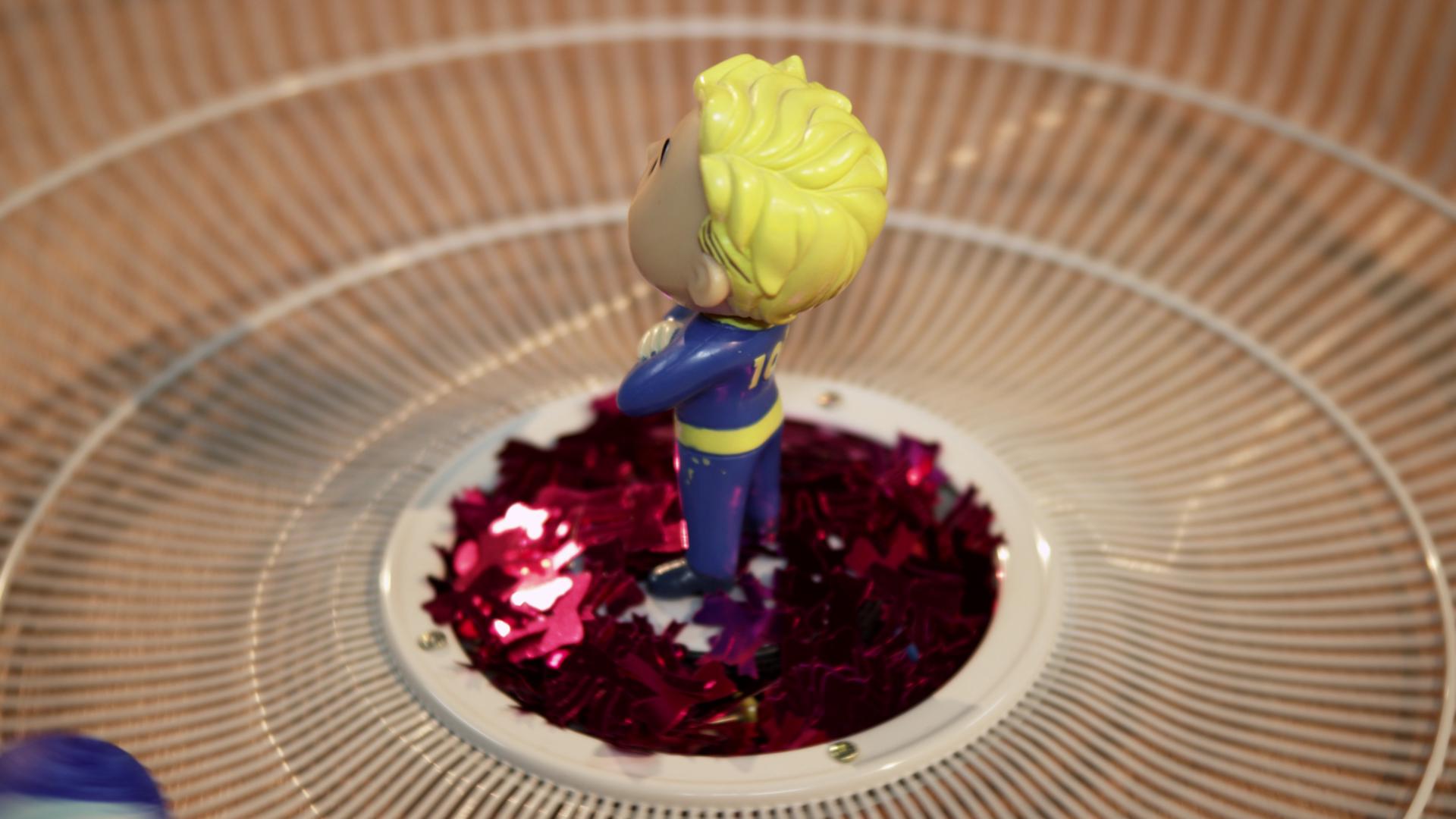}}\;\!\!
    \subfloat[Books]{\includegraphics[width=0.160\linewidth]{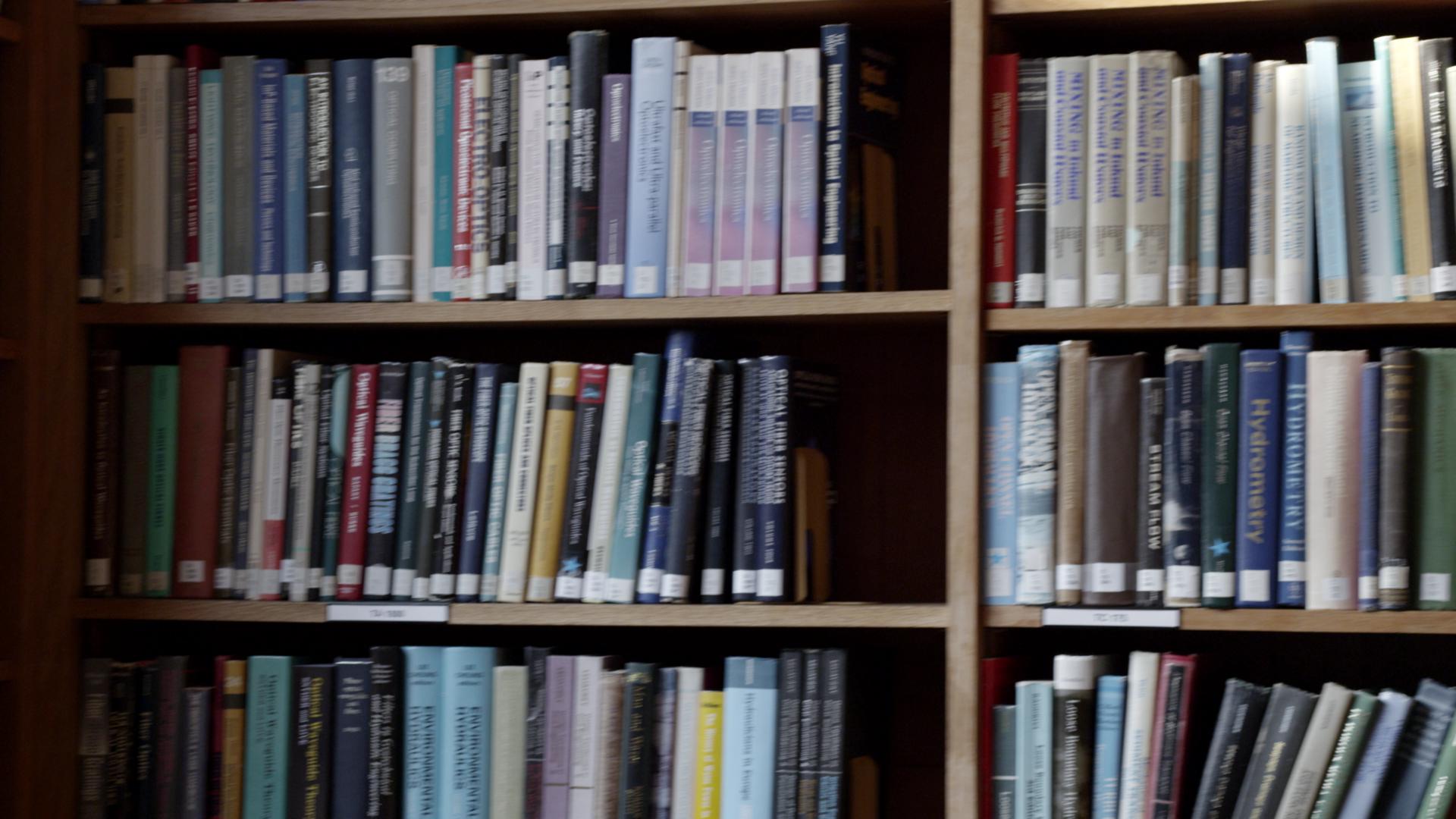}}\;\!\!
    \subfloat[Bouncyball]{\includegraphics[width=0.160\linewidth]{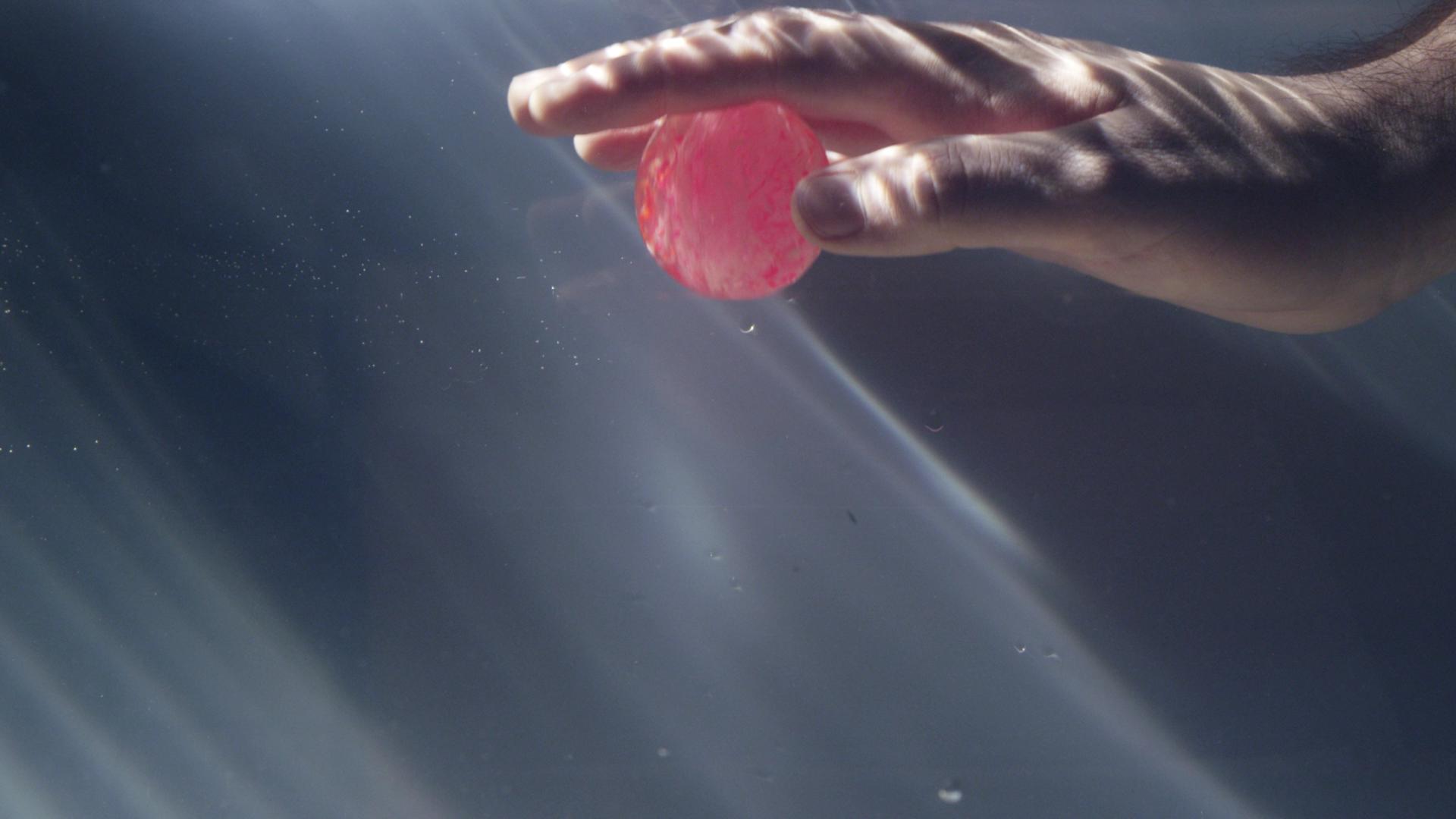}}\;\!\!
    \subfloat[Catch\_track]{\includegraphics[width=0.160\linewidth]{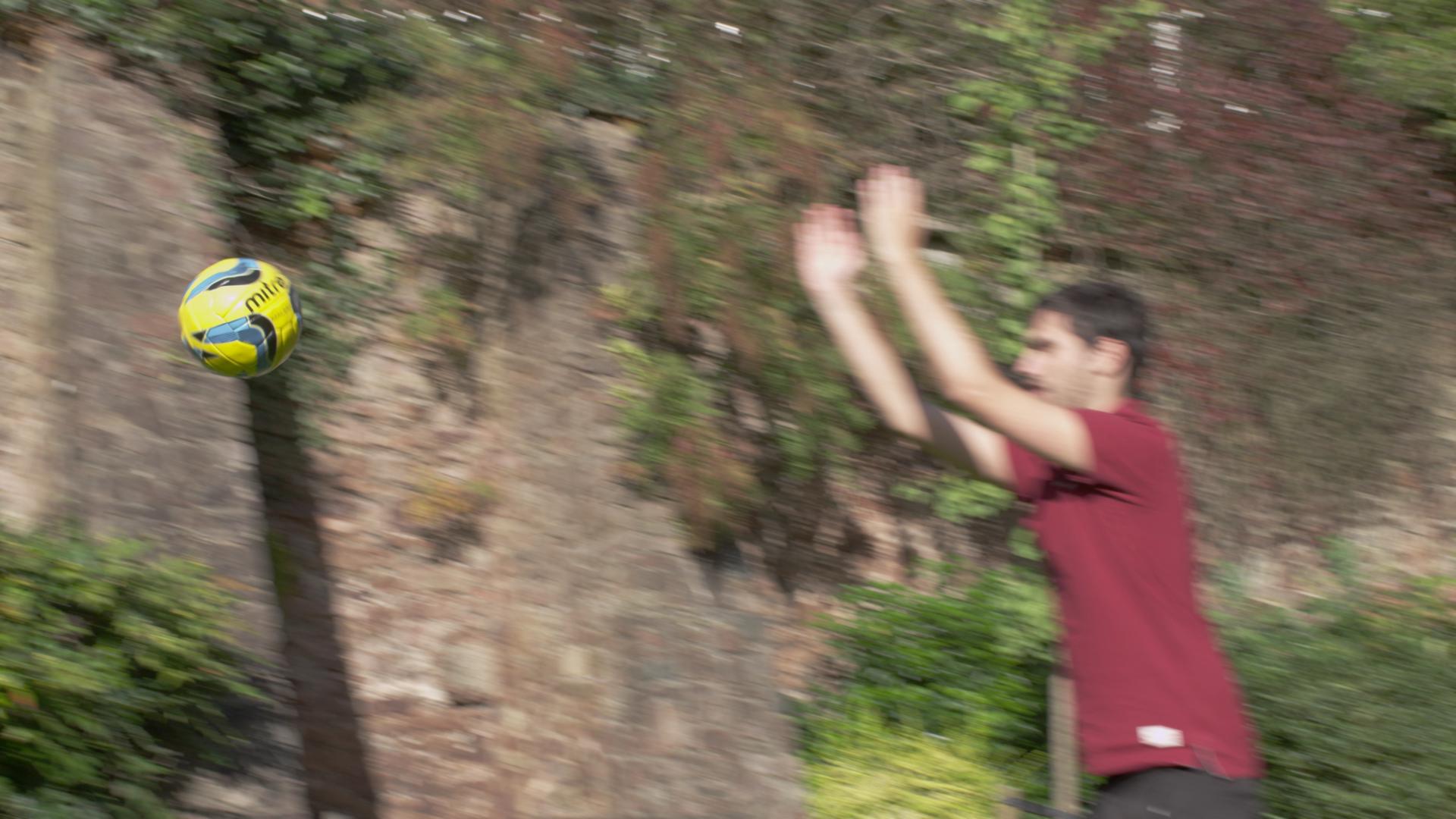}}\;\!\!
    \subfloat[Cyclist]{\includegraphics[width=0.160\linewidth]{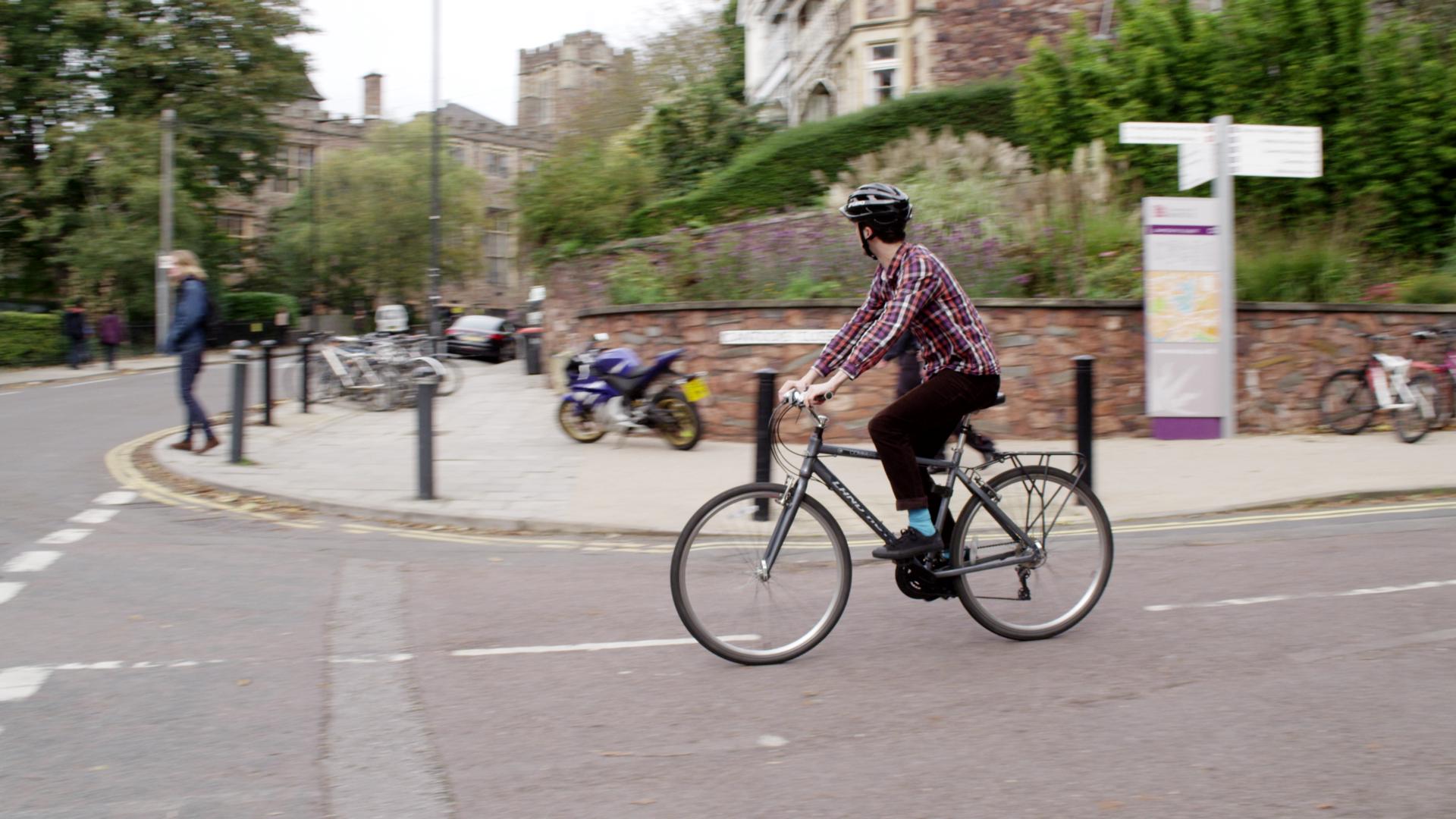}}\;\!\!
    \subfloat[Golf\_side]{\includegraphics[width=0.160\linewidth]{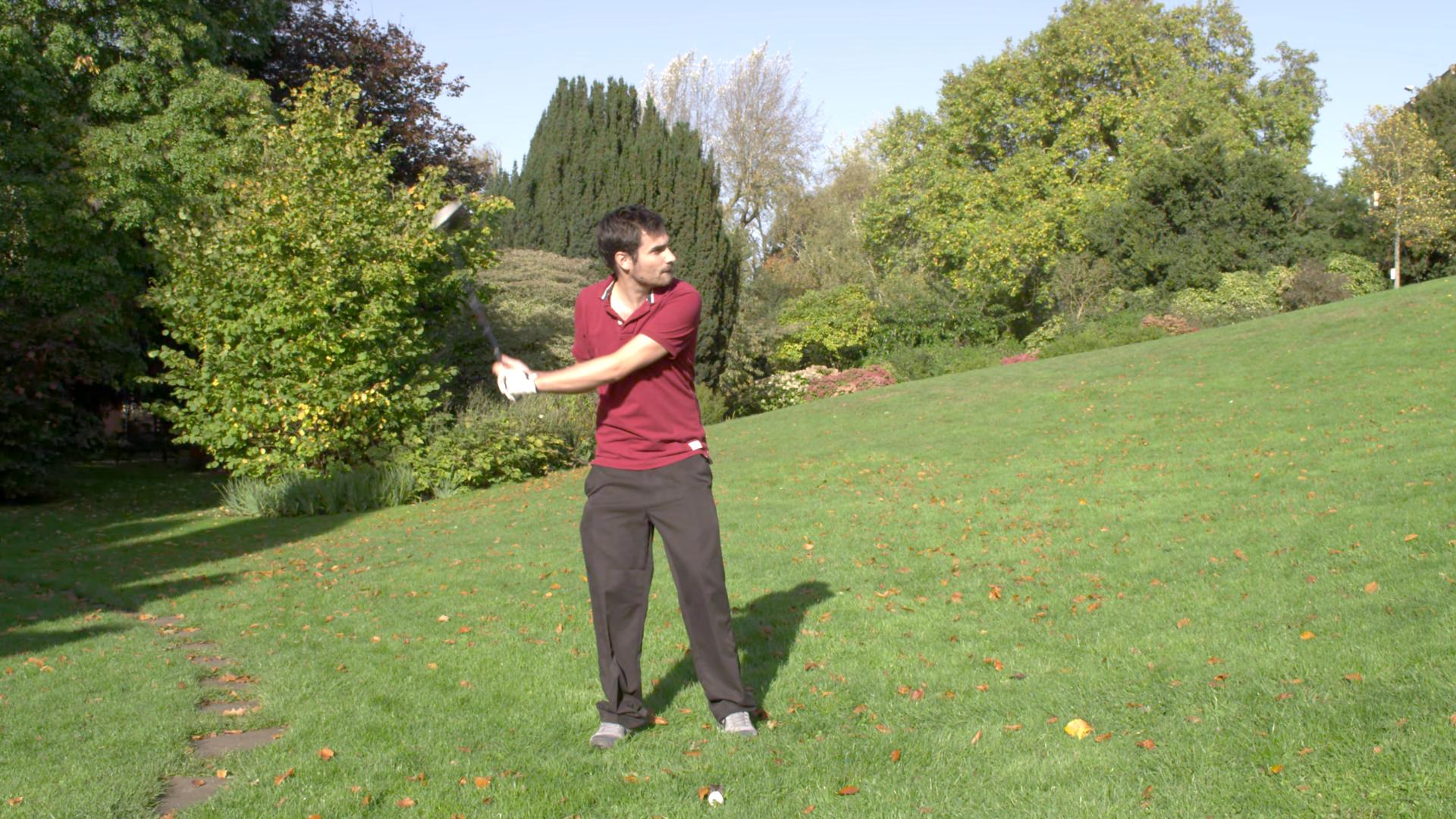}}\\
    \vspace{-3mm}
    \subfloat[Hamster]{\includegraphics[width=0.160\linewidth]{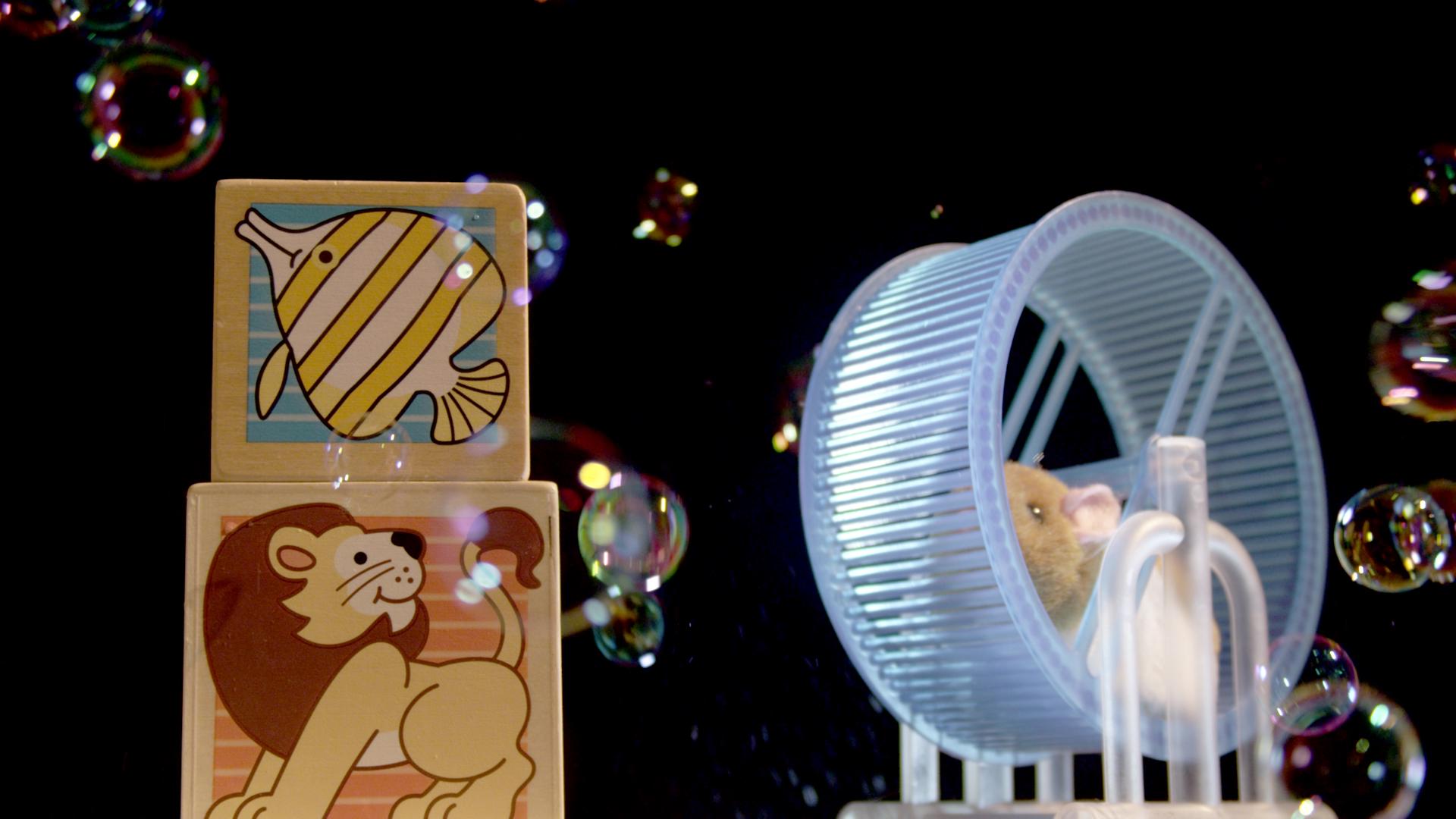}}\;\!\!
    \subfloat[Lamppost]{\includegraphics[width=0.160\linewidth]{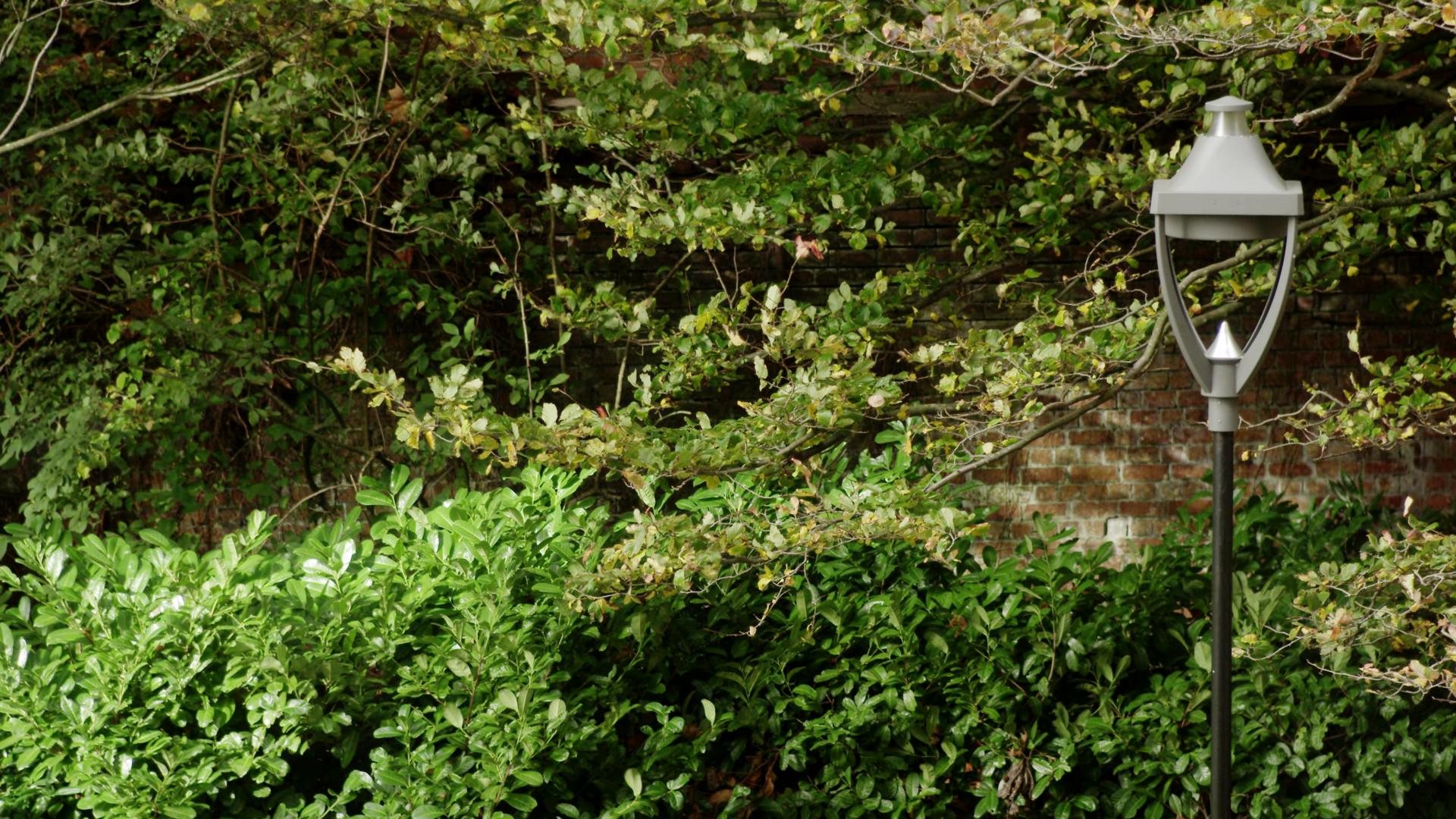}}\;\!\!
    \subfloat[Plasma]{\includegraphics[width=0.160\linewidth]{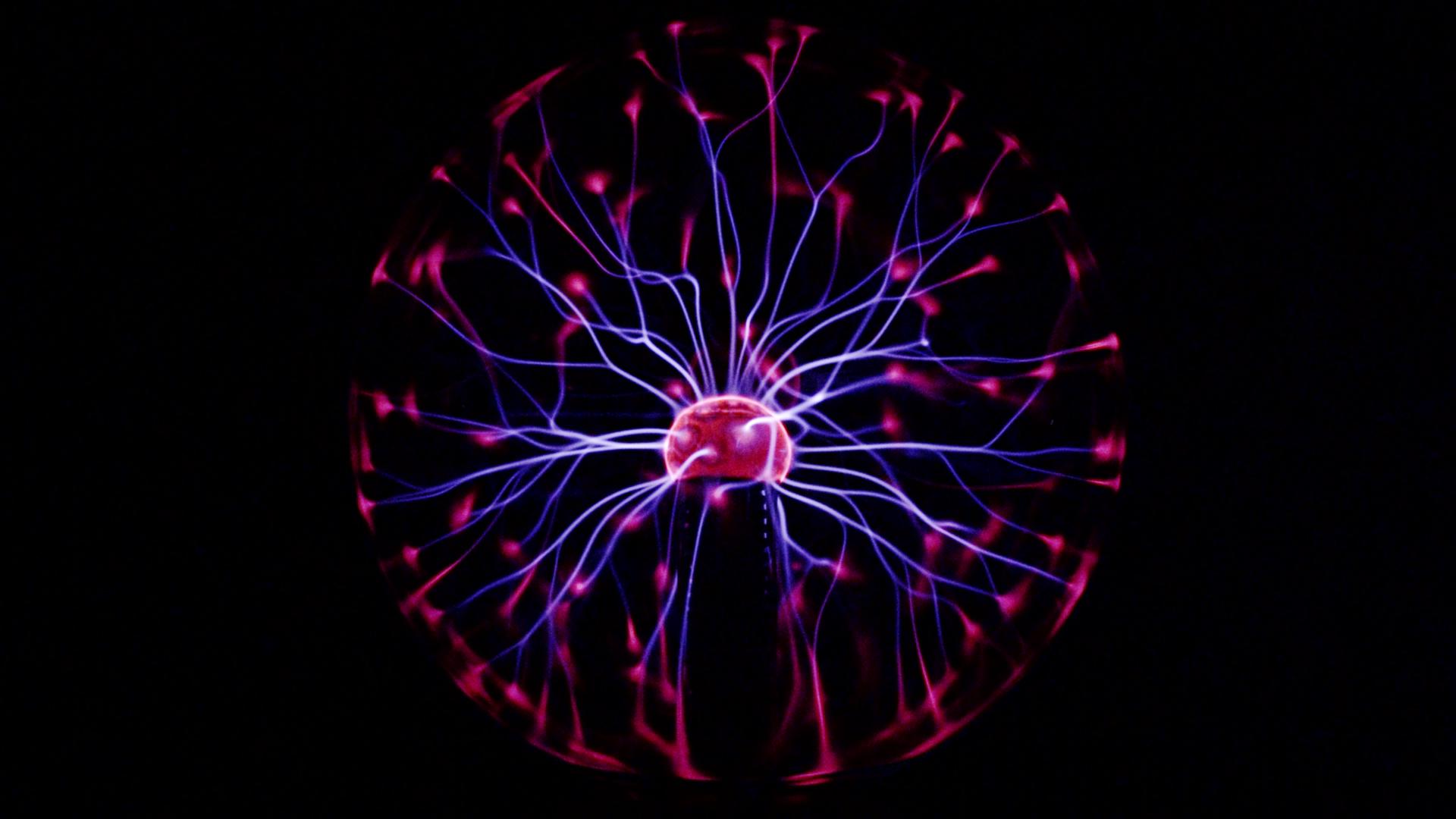}}\;\!\!
    \subfloat[Pond]{\includegraphics[width=0.160\linewidth]{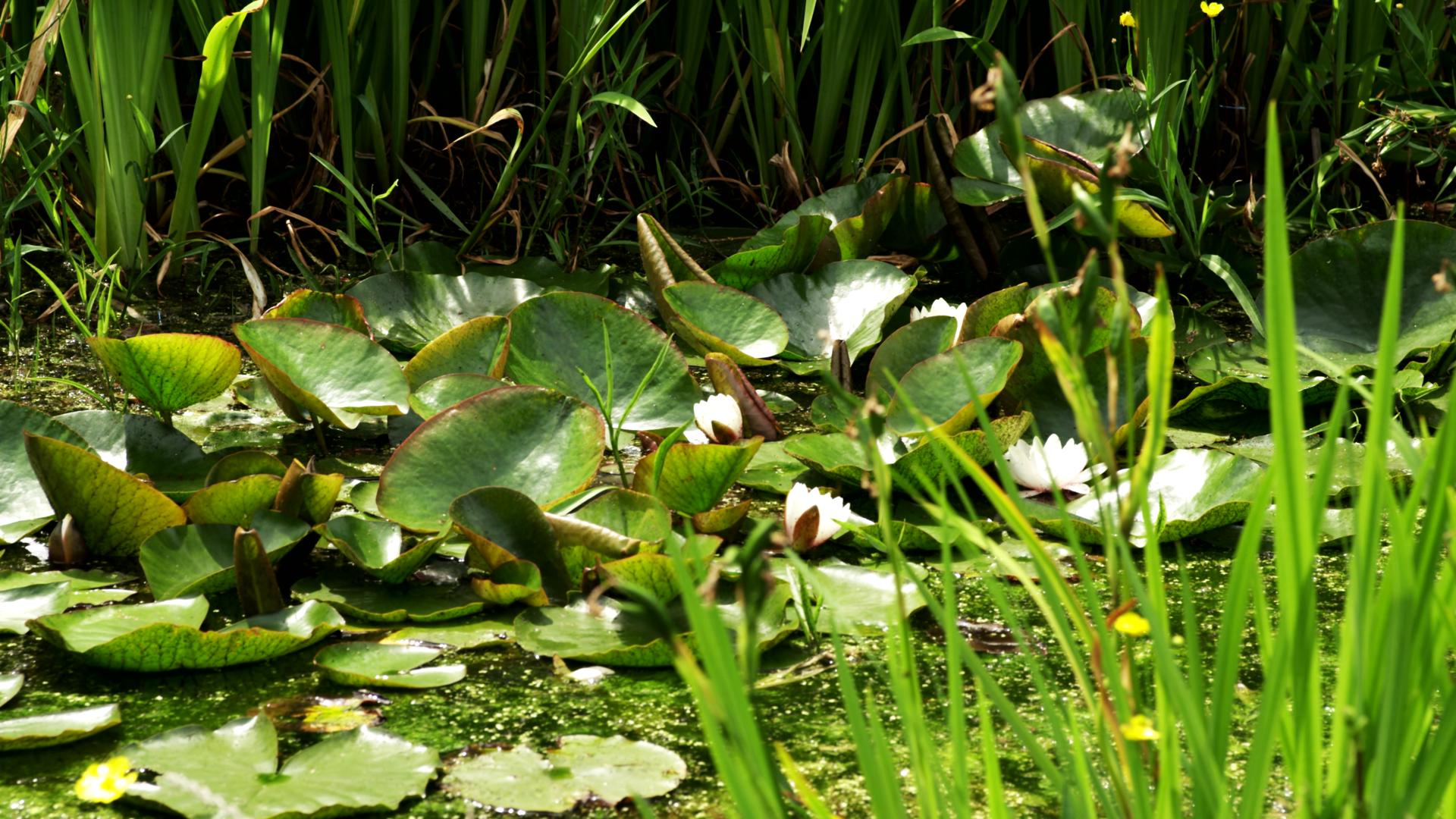}}\;\!\!
    \subfloat[Sparkler]{\includegraphics[width=0.160\linewidth]{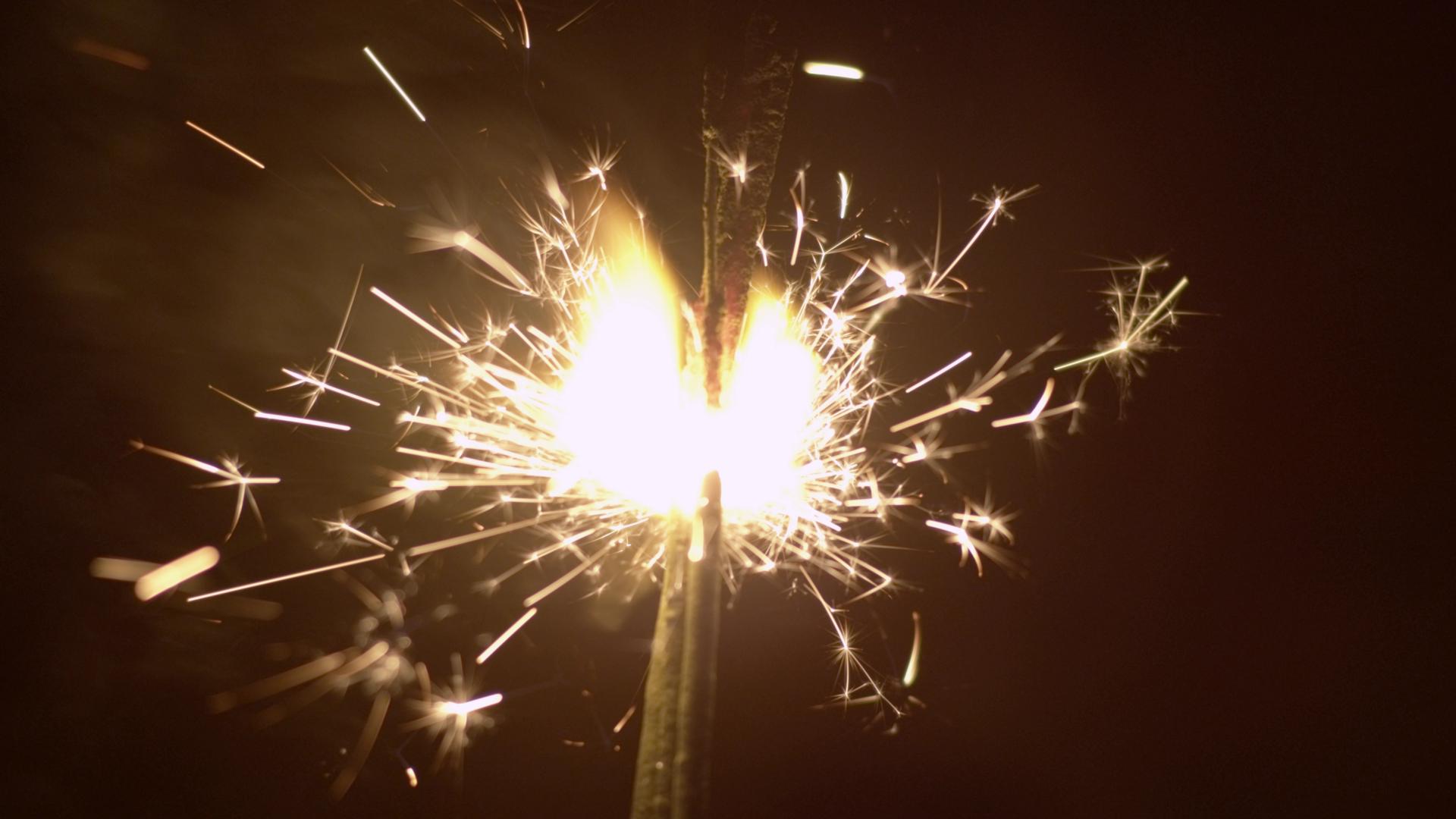}}\;\!\!
    \subfloat[Water\_splashing]{\includegraphics[width=0.160\linewidth]{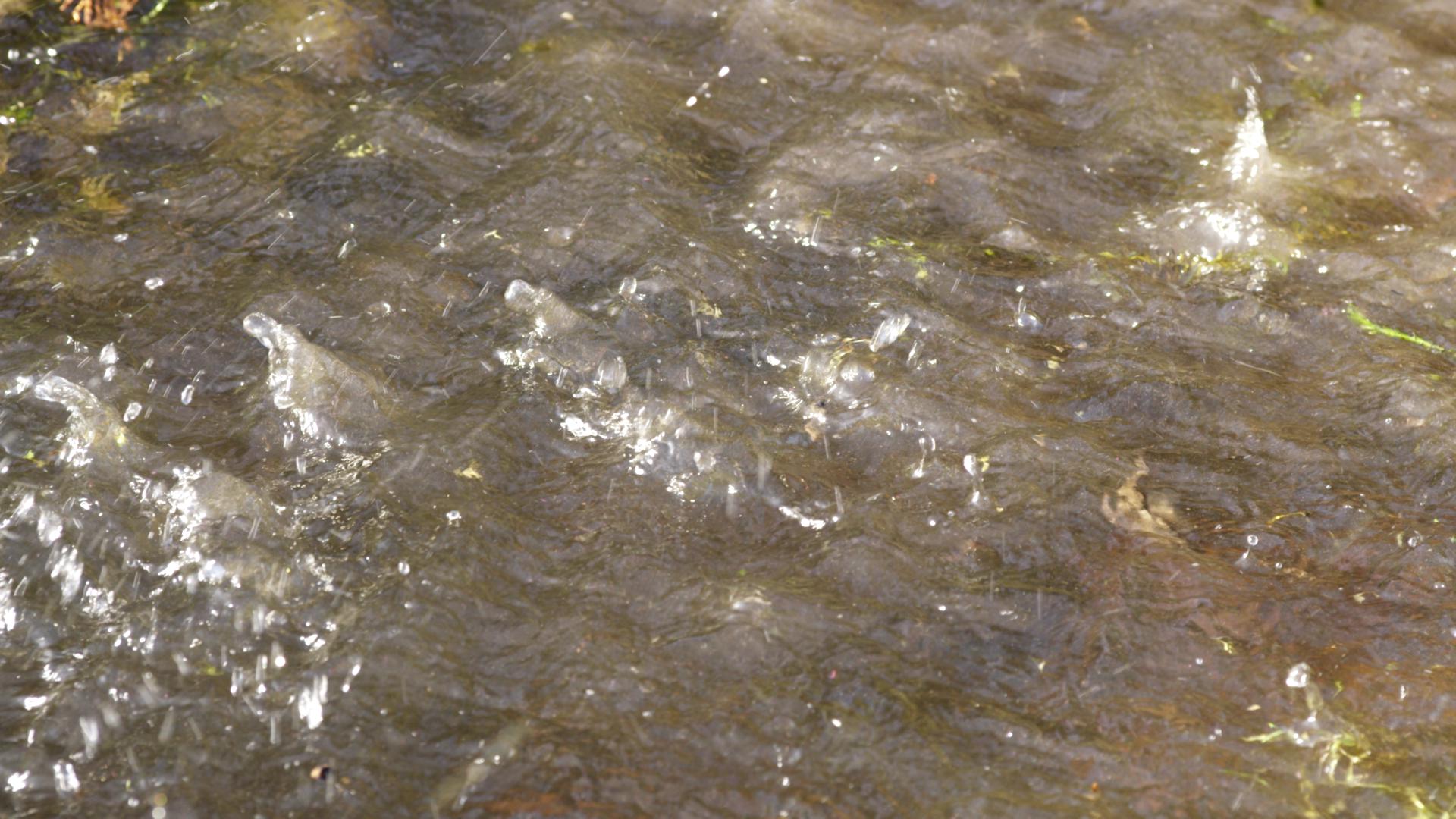}}
    \vspace{-2mm}
    \caption{Sample frames from the reference sequences in BVI-VFI database.}
	\label{fig:example}
\end{figure*}

In order to ensure wide coverage and high uniformity~\cite{winkler2012analysis} across the database, we follow the sequence selection procedure described in \cite{zhang2018bvi} to select 12 source sequences from the initial sequence pool\footnote{This number is set to limit the test time for each subject within 30 min.}. Specifically, we computed four features for each sequence: spatial information (SI), temporal information (TI), motion vector (MV) and dynamic texture parameter (DTP). The latter two were included because motion magnitude and complexity have direct impact on VFI~\cite{danier2021spatio}. The calculation of SI, TI and DTP can be found in \cite{moss2015optimal} and MV is described in \cite{winkler2012analysis}. The sample frames of the final 12 source sequences shown in Fig.~\ref{fig:example}.

The uniformity and coverage of features of the selected source sequences are reported in Table~\ref{tab:feature}, where high range of coverage and excellent uniformity are achieved for all the calculated features when compared to the statistics in~\cite{winkler2012analysis}.

The two lower frame rate versions (60fps and 30fps) of these 12 sources were also included as reference sequences in this database to generate interpolated content at various difficulty levels. This makes a total of 36 reference sequences.

\begin{figure}[t]
    \centering
    \vspace{-3mm}
    \subfloat[Overlay]{\includegraphics[width=0.495\linewidth]{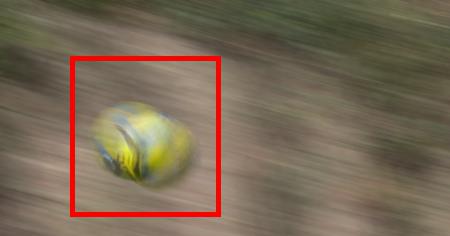}}\;\!\!
    \subfloat[Original]{\includegraphics[width=0.244\linewidth]{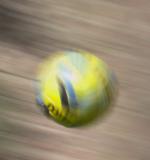}}\;\!\!
    \subfloat[Repeat]{\includegraphics[width=0.244\linewidth]{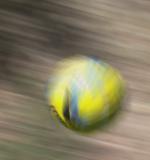}}\\
    \vspace{-3mm}
    \subfloat[DVF]{\includegraphics[width=0.244\linewidth]{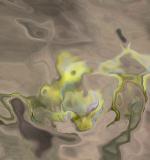}}\;\!\!
    \subfloat[QVI]{\includegraphics[width=0.244\linewidth]{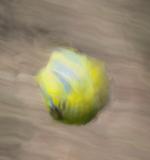}}\;\!\!
    \subfloat[ST-MFNet]{\includegraphics[width=0.244\linewidth]{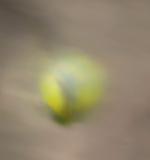}}\;\!\!
    \subfloat[Average]{\includegraphics[width=0.244\linewidth]{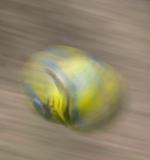}}
    \vspace{-1mm}
    \caption{Example blocks generated by various VFI algorithms. It should be noted that for frame repeating, although the result seems less distorted, the video exhibits motion juddering.}
	\label{fig:distortions}
	\vspace{-4mm}
\end{figure}

\subsection{Test Sequence Generation}

\begin{table}[t]
\centering
\caption{The uniformity and range of coverage for the references sequences in the BVI-VFI dataset.}
\label{tab:feature}
\begin{tabular}{c|c|c|c|c}
\toprule
Feature & SI & TI & DTP & MV \\
\midrule 
Uniformity & 0.943 & 0.970 & 0.916 & 0.767 \\
Range & 0.843 & 0.954 & 0.992 & 0.967\\
\bottomrule	
\end{tabular}
\vspace{-2mm}
\end{table}

To generate different distorted versions, the 36 reference sequences were first down-sampled temporally by a factor of two through frame dropping~\cite{choi2020channel}. The dropped frames were then reconstructed using five different VFI methods using their neighbouring frames. These include frame repeating, frame averaging (from two adjacent frames), DVF~\cite{liu2017video}, QVI~\cite{xu2019quadratic} and ST-MFNet~\cite{danier2021spatio}. The first two were included because they have very low computational complexity and produce unique artifact types, juddering and blurring respectively. The other three methods were all deep learning based approaches. While DVF assumes linear motions between frames, QVI adopts a second order motion model. ST-MFNet is a more recent VFI method using multi-flow based warping~\cite{lee2020adacof}, which enables more complex pixel transformation, and it offers robust interpolation performance for different types of video content. For the three DL based methods, their model parameters trained in \cite{danier2021texture} have been used due to their enhanced performance on challenging content, such as large motion and dynamic textures. As a result, a total of 180 (36$\times$5) distorted videos were generated. Example blocks generated by various VFI methods are shown in Fig.~\ref{fig:distortions}.

\section{Subjective Assessment Experiment}\label{sec:experiment}

This section describes the setup and procedures of the conducted subjective experiment.

\subsection{Experiment Setup}
The psychophysical experiment was conducted in a darkened lab-based environment~\cite{itu2002500}. A BENQ XL2720Z high frame rate monitor with a screen size of 598$\times$336 mm was used to display the sequences. The display resolutions were configured to 1920x1080 spatially, and all the sequences were played at their original spatial resolution and frame rates. The viewing distance was set to 1008 mm (three times the screen height, compliant with ITU-R BT.500~\cite{itu2002500}). The monitor was connected a Windows PC, and Matlab Psychtoolbox 3.0~\cite{pychotoolbox} was used to control this experiment. Subjective scores were collected using a wireless mouse provided to the participants. 

\subsection{Experimental Procedure}
The experiment employed the Double Stimulus Continuous Quality Scale (DSCQS) methodology~\cite{itu2002500}. In each trial, the subject is shown two sequences, A and B, one of which is the distorted video generated by one of the VFI methods, and the other is the corresponding reference with original frames. Their display order in the trial is randomised. After viewing each sequence twice, the participant is presented with a grey screen showing the question: ``Please rate the perceived quality of the video." Two sliders with five evenly spaced ticks labelled \textit{Bad}, \textit{Poor}, \textit{Fair}, \textit{Good}, \textit{Excellent}, corresponding to 0, 25, 50, 75, 100 respectively, are also shown on the screen to record the user input. The user is not told which sequence in this pair is the reference throughout the experiment.

A total of 60 subjects were paid to participate in this experiment, including 32 females and 28 males, with an average age of 25. In order to avoid excessively long time for each test session while ensuring sufficient raw subjective scores for each trial, we followed the approach in \cite{zhang2018bvi} and divided all the participants into three groups. The first group was presented with the material associated with the first four source sequences (1-4 as shown in Fig.~\ref{fig:example}); the second group was assigned the next four source sequences (5-8); the last group was shown the videos from the last four source sequences (9-12). This results in 60 trials (4 source$\times$3 frame rates$\times$5 VFI methods) for each session (participant), and the trial order in each session is also randomised. At the beginning of each session, the participant's visual acuity and colour blindness were assessed using a Snellen chart and a Ishihara chart respectively. S/he was then given instructions and familiarised with four practice trials, which contained different sequences from those in BVI-VFI. Each formal test session took 30 minutes on average.

\begin{figure}[t]
    \centering
    \includegraphics[width=.8\linewidth]{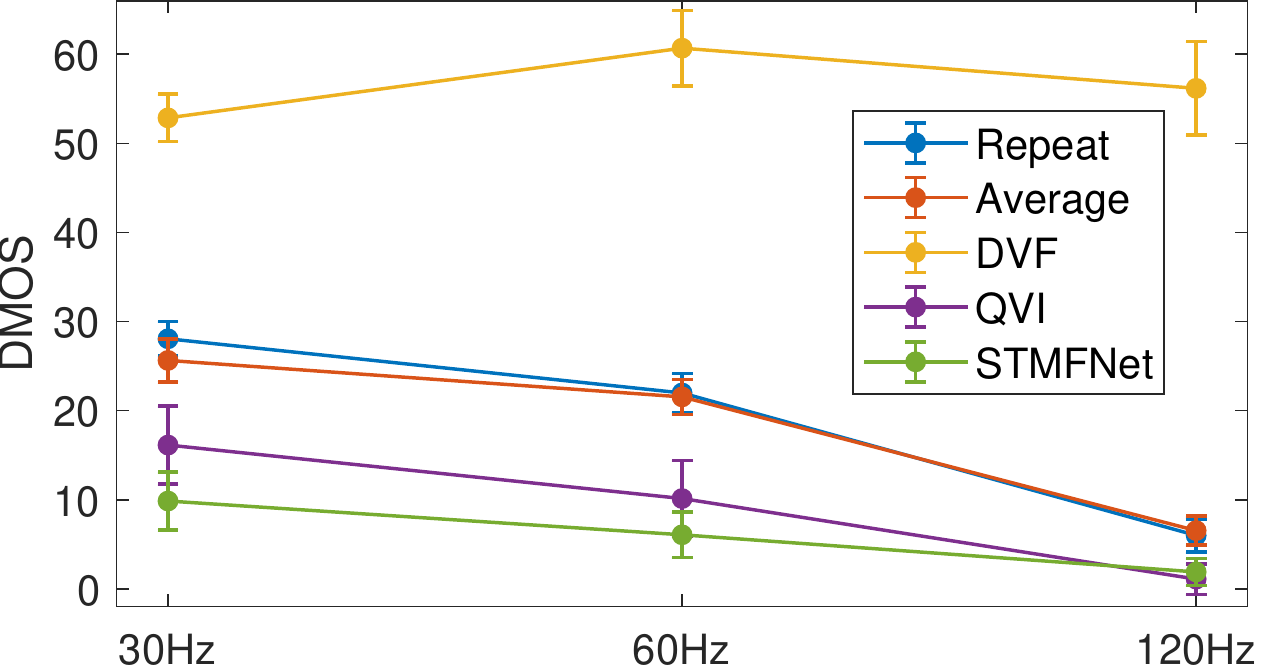}
    \caption{The DMOS values for 5 VFI methods at 3 frame rates. The error bar denotes the standard error over sequences. Note that lower DMOS values correspond to better visual quality.}
	\label{fig:userdata}
	\vspace{-3mm}
\end{figure}

\subsection{Data Processing}
The collected subjective data (i.e. ranging from 0-100) were processed as follows. Firstly, for each subject and each trial, a differential score was obtained by subtracting the score assigned to the distorted video from the score given to the reference video. Then the differential mean opinion score (DMOS) for each distorted sequence was obtained by taking the average of all the differential scores for that sequence. This totalled to 180 DMOS values corresponding to the 180 distorted videos.

\begin{figure*}[t]
    \centering
    \subfloat[]{\includegraphics[width=0.240\linewidth]{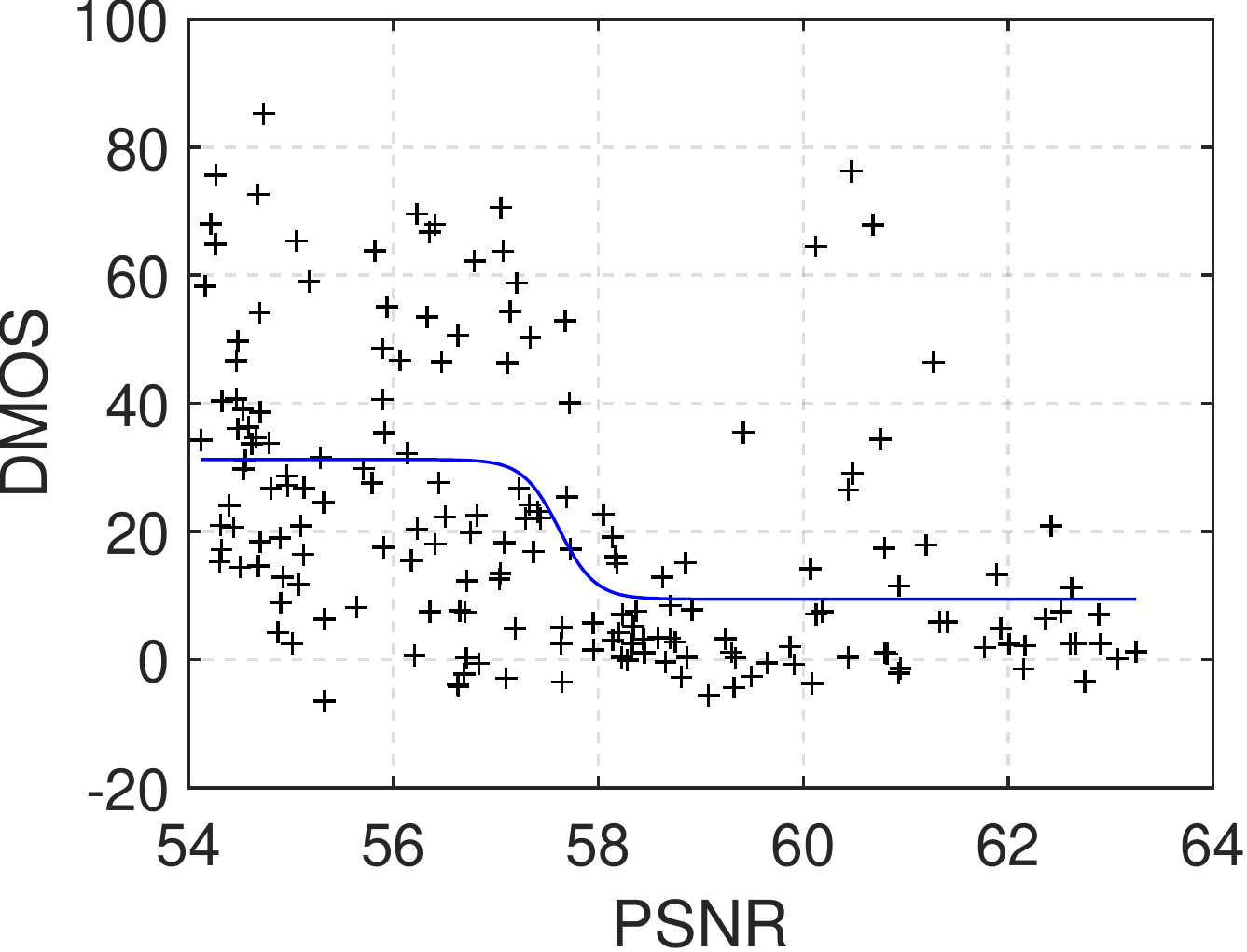}}\;\!\!
    \subfloat[]{\includegraphics[width=0.240\linewidth]{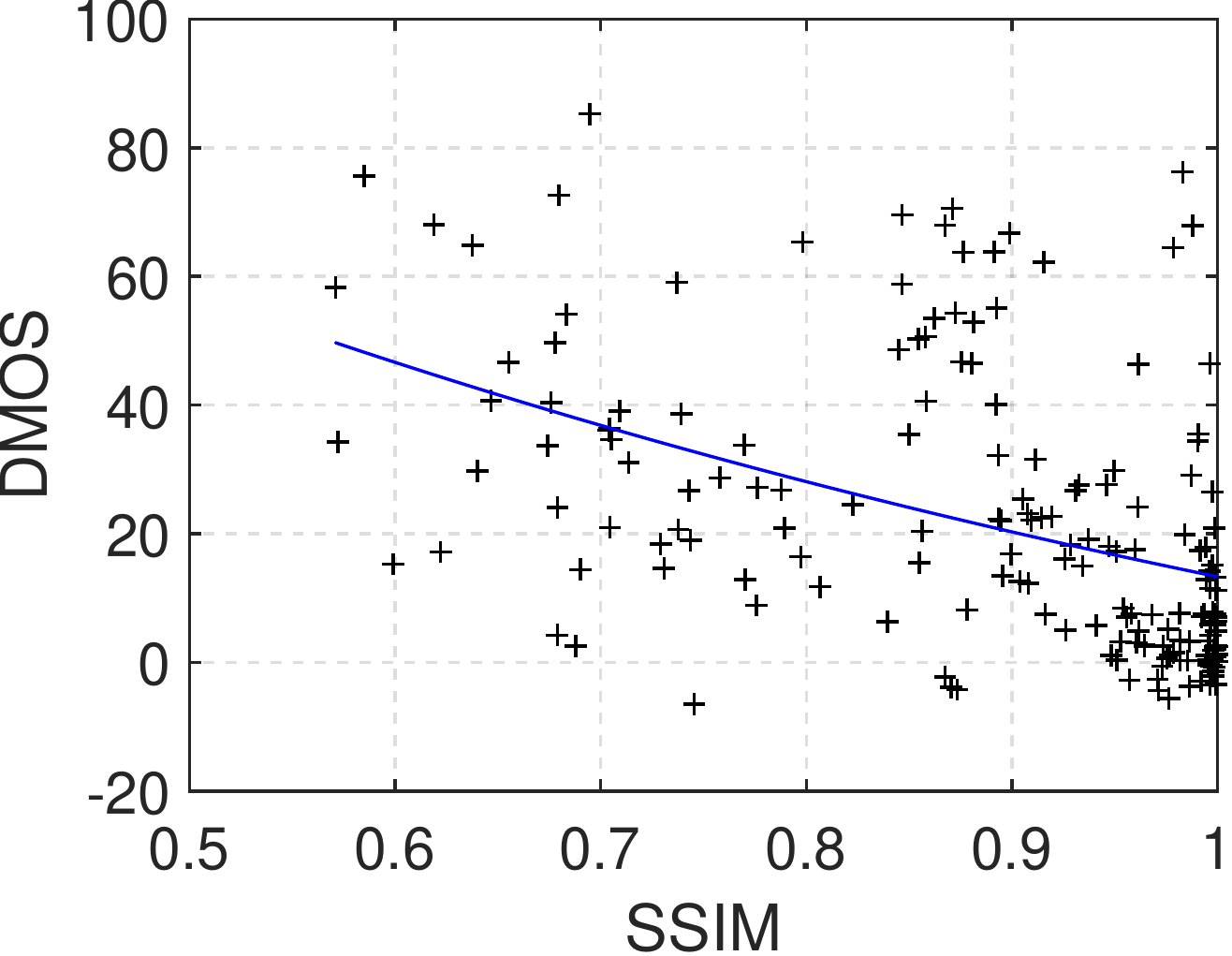}}\;\!\!
    \subfloat[]{\includegraphics[width=0.240\linewidth]{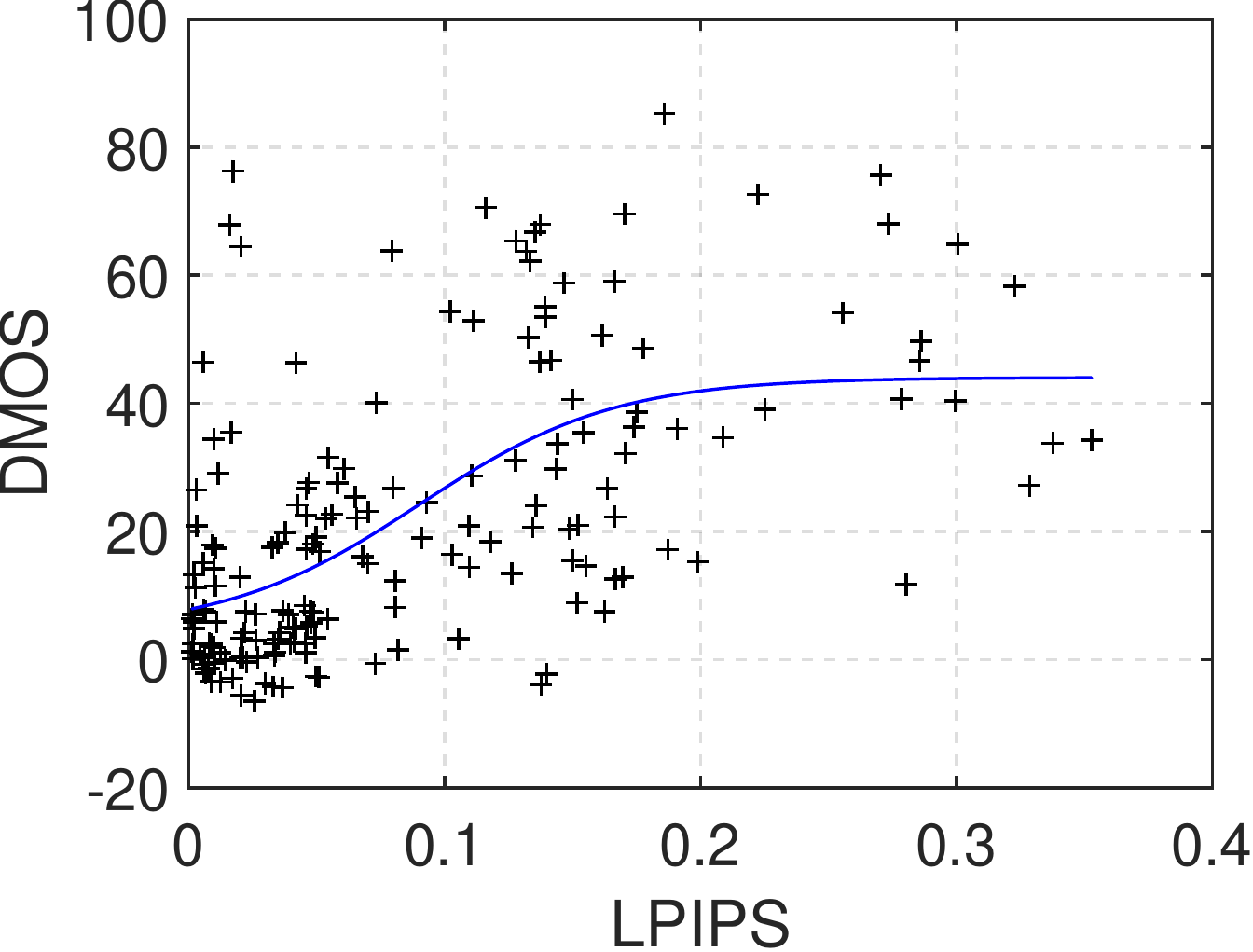}}\;\!\!
    \subfloat[]{\includegraphics[width=0.240\linewidth]{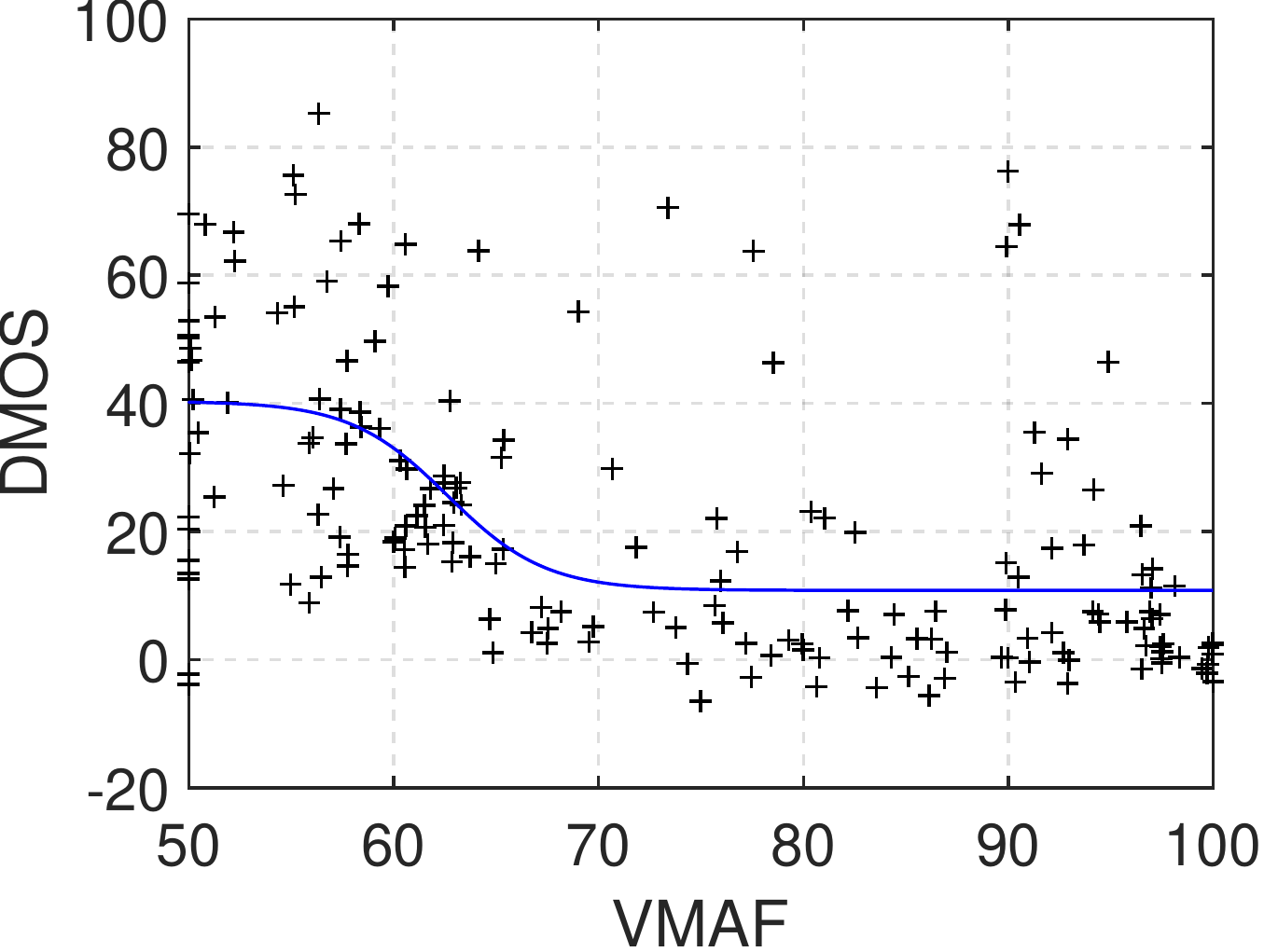}}\\
    \vspace{-2mm}
    \caption{Plots of DMOS values against the scores of selected quality metrics. PSNR, SSIM and LPIPS are commonly used in VFI, and VMAF is the second best performing metric. The blue lines are the logistic function fitted on the BVI-VFI database.}
	\label{fig:metric}
	\vspace{-3mm}
\end{figure*}


\section{Results and Discussion}\label{sec:results}

The DMOS values obtained for each of the five VFI methods at 3 frame rates are plotted in Fig.~\ref{fig:userdata}, where it can be seen that ST-MFNet shows the best overall performance across different frame rates, while QVI performs similarly well at 120fps. It is also observed that content interpolated by DVF received the highest DMOS values for all frame rates, and this may be due to the linear motion assumption made in DVF -- the optical flows between the non-existent middle frame to the next and previous frames are assumed symmetric. This can largely limit the performance of DVF, especially on the sequences in BVI-VFI dataset which contain complex motions. 

\begin{table}[t]
\centering
\caption{The performance of the tested quality assessment models on the BVI-VFI dataset.}
\vspace{-1mm}
\label{tab:metric}
\resizebox{\linewidth}{!}{
\begin{tabular}{c|c|c|c|c|c|c}
\toprule
       & \multicolumn{4}{c|}{All data} & non-DL & DL\\
\midrule
Metric & PLCC & SROCC & OR & RMSE & SROCC & SROCC  \\
\midrule 
PSNR & 0.471 & 0.520 & 0.028 & 19.358 & 0.332 & 0.546  \\
SSIM & 0.475 & 0.581 & 0.044 & 19.328 & 0.305 & 0.627  \\
MSSSIM & 0.529 & 0.593 & 0.033 & 18.623 & 0.334 & \textbf{0.636} \\
LPIPS & \textbf{0.597} & \textbf{0.599} & \textbf{0.022} & \textbf{17.603} & 0.401 & 0.600 \\
VIF & 0.489 & 0.535 & 0.039 & 19.152 & 0.332 & 0.548 \\
FRQM & 0.456 & 0.535 & 0.033 & 19.525 & \textbf{0.725} & 0.457 \\
ST-GREED & 0.214 & 0.112 & 0.050 & 21.432 & 0.064 & 0.152 \\
VMAF & 0.564 & 0.595 & 0.039 & 18.115 & 0.345 & 0.628\\
\bottomrule	
\end{tabular}
}
\vspace{-4mm}
\end{table}

These DMOS values were then used to evaluate their correlation with eight quality metrics:  PSNR, SSIM~\cite{wang2004image}, MSSSIM~\cite{wang2003multiscale}, LPIPS~\cite{zhang2018unreasonable}, VIF~\cite{sheikh2005information}, VMAF~\cite{li2016toward}, FRQM~\cite{zhang2017frame} and ST-GREED~\cite{madhusudana2021st}. While PSNR, SSIM, MSSSIM, and LPIPS are commonly used for assessing video frame interpolation methods, VIF and VMAF are included due to their superior correlation with perceptual quality in other application scenarios (e.g. video compression~\cite{zhang2018bvi}). Additionally, FRQM is designed to measure video quality when frame rate is altered (reduced), which is relevant to VFI. Finally, ST-GREED is a machine learning based approach that concerns both compression and temporal down-sampling artefacts.

The quality metrics were evaluated based on four statistical parameters, Pearson Linear Correlation Coefficient (PLCC), Spearman Rank Order Correlation Coefficient (SROCC), Outlier Ratio (OR) and Root Mean Squared Error (RMSE) \cite{zhang2021intelligent}. For the calculation of PLCC, OR and RMSE, a logistic function is fit on the ground-truth DMOS values and the computed metric scores as described in \cite{video2000final}. 

The performance of all eight evaluated quality metrics is summarised in Table~\ref{tab:metric}, where it can be observed that, for the whole BVI-VFI database, the best performing metric, LPIPS, revealed PLCC and SROCC values lower than 0.6. We have further subgrouped the database into two classes based on the nature of the interpolation methods. For content interpolated by non deep learning approaches, FRQM offers the highest SROCC of 0.725. MSSSIM appears to be the top metric for evaluating test sequences interpolated by deep learning (DL) based VFI methods. The DMOS and metric values and the fitted logistic curves for PSNR, SSIM, LPIPS and VMAF are also plotted in Fig.~\ref{fig:metric}. 

After manual inspection of the sequences whose quality indices are significant outliers away from the fitting curves, we found that most tested quality metrics tend to fail when the scene contains a static background behind fast-moving objects, e.g. the sequences Golf\_side and Hamster. In these cases, all the VFI methods are likely to produce non-uniformly distributed artefacts, primarily around the small areas with foreground moving objects. This leads to poor assessment performance of these metrics due to their employed spatial pooling method, arithmetic mean.

To further assess the statistical significance of the metric performance difference, for every pair of metrics, an F-test is performed on the residuals between the true DMOS values and the DMOS predicted by the metrics using the fitted logistic function~\cite{seshadrinathan2010study}. With 95\% confidence, most metrics are equivalent based on the F-test, except that MSSSIM, LPIPS and VMAF are superior to ST-GREED.

In summary, our results indicate that none of the metrics evaluated here (including PSNR, SSIM and LPIPS that are most commonly employed for VFI) offer satisfactory correlation with perceptual quality on temporally interpolated videos. This means that there is an urgent need to develop a bespoke perceptual quality metric for video frame interpolation, especially for the evaluation of DL based VFI methods, where the highest SROCC achieved is 0.636 (by MSSSIM).

\section{Conclusion}\label{sec:conclusion}

This paper presents subjective study results based on a novel video quality database, BVI-VFI, dedicated to video frame interpolation. This database consists of 36 reference sequences with diverse motion types and their corresponding 180 distorted sequences. The subjective study was conducted to collect opinion scores on the distorted sequences, and the results were then used to evaluate eight quality metrics commonly used in VFI and other application scenarios. The results indicate that none of the tested metrics offer acceptable correlation performance with perceptual quality on this database. There is thus an urgent need for research into a bespoke perceptual quality metric which offers dramatically improved assessment performance for VFI.

\vfill\pagebreak

\small
\bibliographystyle{IEEEtran}
\bibliography{egbib}

\end{document}